\documentclass[11pt,twoside]{article}

\usepackage{asp2006}
\usepackage{epsf}
\usepackage{graphicx}
\usepackage{lscape}
\usepackage{natbib}

\begin{document}
\title{Blue Dots Team Transits Working Group Review}   

\author{A. Sozzetti\altaffilmark{1}, C. Afonso\altaffilmark{2}, R. Alonso\altaffilmark{3}, 
D. L. Blank\altaffilmark{4}, C. Catala\altaffilmark{5}, H. Deeg\altaffilmark{6}, 
J. L. Grenfell\altaffilmark{7}, C. Hellier\altaffilmark{8}, D. W. Latham\altaffilmark{9}, 
D. Minniti\altaffilmark{10}, F. Pont\altaffilmark{11}, and H. Rauer\altaffilmark{12}}   
\altaffiltext{1}{INAF - Osservatorio Astronomico di Torino, Strada Osservatorio 20, I-10025 Pino Torinese, Italy}
\altaffiltext{2}{Max Planck Institute for Astronomy, K\"onigstuhl 17, 69117 Heidelberg, Germany}
\altaffiltext{3}{Observatoire de Gen\'eve, Universit\'e de Gen\`eve, 51 Ch. des Maillettes, 1290 Sauverny, Switzerland}
\altaffiltext{4}{James Cook University, Townsville, QLD 4811, Australia}
\altaffiltext{5}{LESIA, Observatoire de Paris, 5 place Jules Janssen 92195 Meudon Cedex, France}
\altaffiltext{6}{Instituto de Astrof{\`\i}sica de Canarias, 38205 La Laguna, Tenerife, Spain}
\altaffiltext{7}{Technische Universit\"at Berlin, Hardenbergstr. 36, 10623 Berlin, Germany}
\altaffiltext{8}{Keele University, Staffordshire, ST5 5BG, UK}
\altaffiltext{9}{Harvard-Smithsonian Center for Astrophysics, 60 Garden Street, Cambridge, MA 02138, USA}
\altaffiltext{10}{Pontificia Universidad Cat\'olica de Chile, Casilla 306, Santiago 22, Chile}
\altaffiltext{11}{University of Exeter, Exeter EX4 4QL, UK}
\altaffiltext{12}{Institute of Planetary Research (DLR), Rutherfordstr. 2, 12489 Berlin, Germany}

\begin{abstract} 
Transiting planet systems offer an unique opportunity to observationally constrain
proposed models of the interiors (radius, composition) and atmospheres (chemistry, dynamics)
of extrasolar planets. The spectacular successes of ground-based transit surveys
(more than 60 transiting systems known to-date) and the host of multi-wavelength, spectro-photometric
follow-up studies, carried out in particular by HST and Spitzer, have paved the way to the next
generation of transit search projects, which are currently ongoing (CoRoT, Kepler),
or planned. The possibility of detecting and characterizing transiting 
Earth-sized planets in the habitable zone of their parent stars appears tantalizingly close.
In this contribution we briefly review the power of the transit technique for characterization of
extrasolar planets, summarize the state of the art of both ground-based and space-borne
transit search programs, and illustrate how the science of planetary transits fits within
the Blue Dots perspective.
\end{abstract}


\section{Introduction}   

Within the framework of the Blue-Dots Team (BDT) initiative (http://www.blue-dots.net/),
the primary goal of the Transits Working Group (TWG) is to gauge the potential and
limitations of transit photometry (and follow-up measurements) as a tool to detect
and characterize extrasolar planets, while emphasizing its complementarity with other
techniques. Following the `grid approach' agreed upon within the BDT,
the mapping is to be performed as a function of depth of the science investigation,
project scale, and detectable exoplanet class. In this review of the TWG activities,
we first describe the main observables accessible by means of photometric transits.
We then focus on the host of follow-up techniques that can be utilized to deepen our
understanding and characterization of transiting systems, and briefly touch upon some
key science highlights, while keeping in mind the primary difficulties and limitations
inherent to the transit technique when applied to planet detection.
Next, a summary of the present-day and future projects devoted
to detection and characterization of transiting planets, both from the ground and in
space, is presented. Finally, we use a `grid approach' to properly gauge how
the scientific prospects of photometric transits and follow-up techniques fit
within the BDT perspective.

\section{Planet Detection with Transit Photometry}

\subsection{Observables}

The primary observable of the transit technique is the periodic decrease of
stellar brightness of a target, when a planet moves across the stellar disk. The
magnitude of the eclipse depth is defined as $\Delta F/F_0 = (R_p/R_\star)^2$, where
$F_0$ is the measured out-of-transit flux, and $R_p$ and $R_\star$ are the planet's
and primary radius, respectively. Transits require a stringent condition of observability,
i.e. the planetary orbit must be (almost) perpendicular to the plane of the sky. The
geometric probability of a transit, assuming random orientation of the planetary
orbit, is: $P_{tr} = 0.0045(1\,\mathrm{AU}/a)((R_\star+R_p)/R_\odot)((1+e\cos(\pi/2-\varpi))/(1-e^2))$, 
where $a$ is the orbital semi-major axis, $e$ the eccentricity,
and $\varpi$ the argument of periastron passage (e.g.,~\citealt{charbon07}). 
The duration of a central transit is $\tau =
13(R_\star/R_\odot)(P/1\,\mathrm{yr})^{(1/3)}(M_\odot/M_\star)^{(1/3)}$ hr,
with a reduction for non-central transits by a factor
$\sqrt{1-b^2/R_\star^2}$, where $b = (a/R_\star)\cos i$ is the
impact parameter for orbital inclination $i$ (e.g.,~\citealt{seager03}). The transit method allows to determine
parameters that are not accessible to Doppler spectroscopy, such
as the ratio of radii, the orbital inclination, and the stellar
limb darkening. When combined with available radial-velocity (RV) 
observations, actual mass and radius estimates for the planet can
be derived, provided reasonable guesses for the primary mass and
radius can be obtained.

\subsection{False positives reconnaissance}\label{followup}

The typical transiting system configurations known to-date (for a comprehensive list
see for example http://exoplanet.eu) encompass jovian planets orbiting solar-type stars on
a few-day orbits, resulting in eclipse depths of $0.3-3\%$ and transit durations of $1.5-4$ hr.
A great variety of stellar systems can reproduce such signals in terms of depth and duration.
These include grazing eclipsing binaries, large stars eclipsed by small stars, and `blends'
consisting of faint eclipsing binaries whose light is diluted by a third, brighter star
(e.g.,~\citealt{brown03}). Typically, such impostors constitute over 95\% of the detected transit-like signals in
wide-field photometric surveys datasets. Extensive campaigns of follow-up observations of transit candidates
must then be undertaken in order to ascertain the likely nature of the system. High-quality
light curves and moderate precision ($\approx$ km s$^{-1}$), low signal-to-noise ratio 
(SNR) spectroscopic measurements are usually gathered to
deepen the understanding of the primary, the consistency of the transit shape with that produced
by a planet, and, via detailed modeling of the combined datasets (e.g.,~\citealt{torres04}),
to rule out often subtle blend configurations. Only if the candidate passes all the above tests,
does one resort to use 10-m class telescopes and high-resolution, high-precision Doppler measurements
to determine the actual spectroscopic orbit (e.g.,~\citealt{mandu05}). The plague
of false positives contamination is however rather diminished for ground-based transit surveys targeting 
cool, nearby M-dwarf stars~\citep{nutz08}. This holds true also for 
space-borne transit programs, with very high photometric precision which allows to 
reveal many of the stellar companions via very shallow secondary eclipses, and/or 
ellipsoidal variations out of transit. Giant stars can be excluded beforehand by using a 
target catalog, such as the Kepler Input Catalog or Gaia data (as it's envisioned for PLATO). 
Background eclipsing binaries can be instead identified efficiently if precision astrometry 
can be performed on the photometric times series themselves, to measure centroid shifts 
due to variability-induced movers~\citep{wielen96}. This technique is currently being applied with 
considerable success to Kepler data (Latham et al., this volume). 

\subsection{Transiting Systems Highlights}

\begin{figure}
\centering
\includegraphics[width=.90\textwidth,angle=0.]{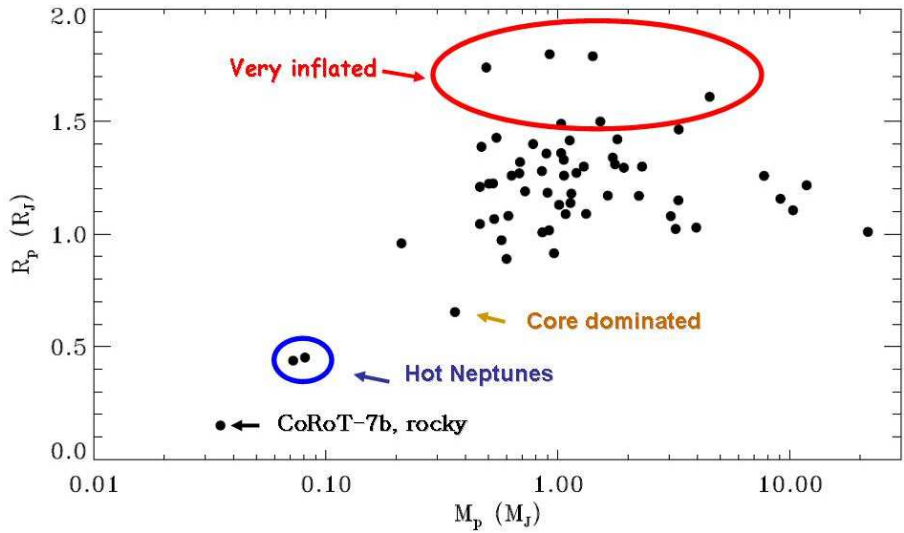}
\caption{Masses and radii for the sample of transiting planets known as of December
2009. }
\label{mrrel}
\end{figure}

Sixty two transiting planets of main-sequence stars are known today. The direct measure
of their masses and radii, and thus densities and surface gravities, puts fundamental
constraints on proposed models of their physical structure (\citealt{charbon07},
and references therein). Figure~\ref{mrrel} shows the $M_p$-$R_p$ relation for the
known transiting systems. Strongly irradiated planets cover a range of
almost three orders of magnitude in mass and more than one order of magnitude in radius.
The variety of inferred structural properties is posing a great challenge to evolutionary
models of their interiors (e.g.,~\citealt{baraffe08,valencia07,miller09}). Among the
most interesting systems found by photometric transit surveys, are $a)$ those containing super-massive
($M_p \approx 7-13$ $M_J$) hot Jupiters such as HAT-P-2b~\citep{bakos07}, WASP-14b~\citep{joshi09}, 
XO-3b~\citep{johns08}, and particularly WASP-18b~\citep{hellier09},
the first planet to be likely tidally disrupted on a short timescale,
$b)$ very inflated ($\sim 1.8$ $R_J$) jovian planets such as
WASP-12b~\citep{hebb09}, TrES-4~\citep{mandu07}, and WASP-17b~\citep{anders09}, 
$c)$ the tilted, most eccentric planet, HD 80606b
(\citealt{winn09}, and references therein),
$d)$ the first transiting planet in a multiple system, HAT-P-13b~\citep{bakos09}, 
$e)$ the first transiting low-mass brown dwarf, CoRot-3b~\citep{deleuil08}, 
and $f)$ the first transiting Super Earth, CoRot-7b~\citep{leger09}, 
itself a member of a multiple-planet system~\citep{queloz09}.

\section{Transiting Planet Characterization: Follow-up Techniques}

When the primary is sufficiently bright (see below), a host of follow-up photometric
and spectroscopic measurements can be carried out over a large range of wavelengths, to
deepen the characterization of the physical and dynamical properties of 
transiting systems~\citep{charbon07}.
At visible wavelengths, long-term, high-cadence, high-precision photometric monitoring
can allow to detect additional components in a system (not necessarily transiting) via the
transit time variation method~\citep{holman05}, while RV measurements collected
during transit offer the opportunity to determine the degree of alignment between the stellar
spin and the orbital axis of the planet (e.g.,~\citealt{winn09}). These data
are powerful diagnostics for models of orbital migration and tidal evolution of
planetary systems (e.g.,~\citealt{fabr09}, and references therein).
The technique of transmission spectroscopy
opens the way to measurements of specific elements seen in absorption in the planet's atmosphere
~\citep{charbon02}, including water~\citep{tinetti07}.
At infrared wavelengths, mostly thanks to the Spitzer Space
Telescope, the photometric and spectroscopic monitoring over a wide range of planetary
phases, and particularly during secondary eclipse, has allowed to study in detail the strongly-irradiated
atmospheres of a few planets, with successful detection of the planet's thermal emission
~\citep{charbon05} and characterization of the longitudinal temperature distribution
(e.g.,~\citealt{knut07,charbon07}).
The direct measurements of exoplanets' atmospheric compositions and temperature profiles,
atmospheric dynamics, and phase light curves are key inputs for models of atmospheric physics,
chemistry, and dynamics~\citep{burrows08}.  

\section{Ground-based and space-borne projects}

\begin{figure}
\centering
\includegraphics[width=.62\textwidth,angle=0.]{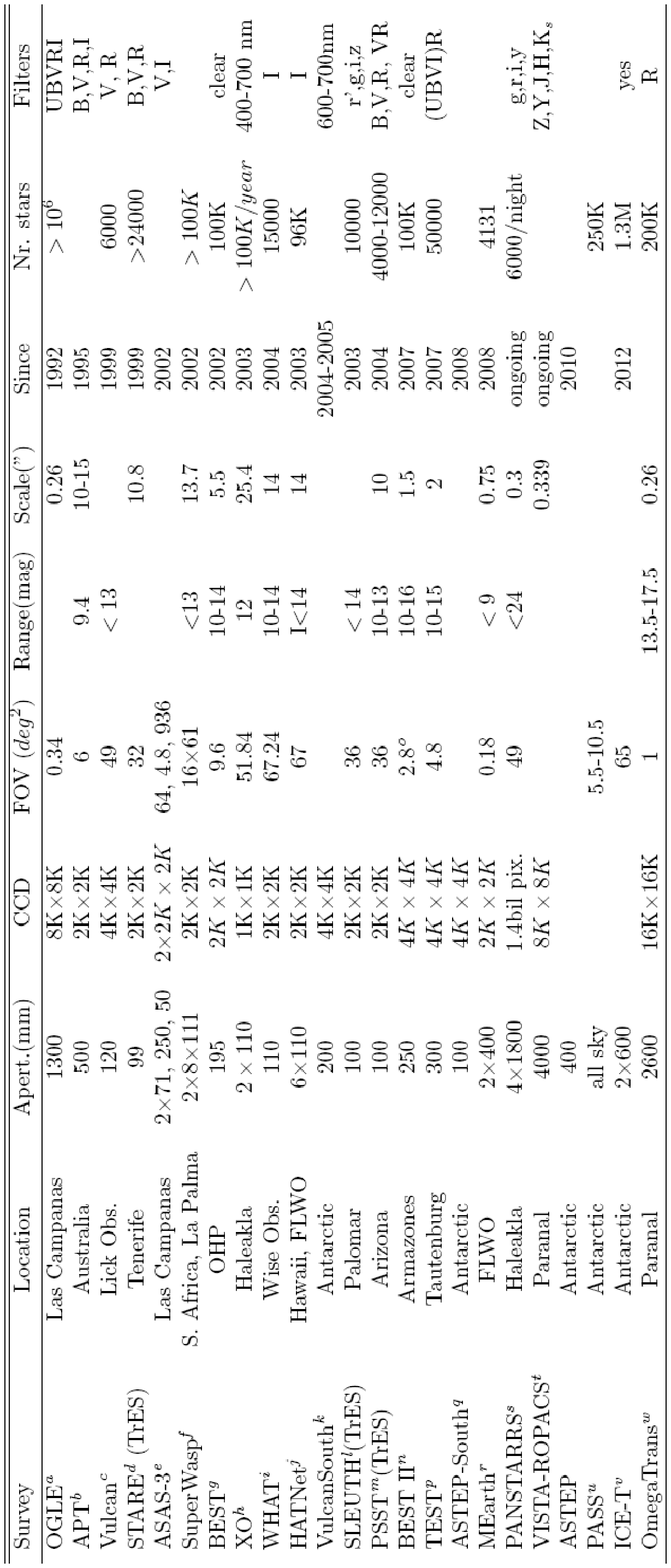}
\caption{Ground-based transit surveys summary table.}
\label{groundtab}
\end{figure}

The success of (wide-field) ground-based photometric transit surveys bears upon two distinct
approaches, the first adopting moderate-size telescopes to search relatively faint stars,
the second utilizing small-size instrumentation for searches around brighter targets
(see Figure~\ref{groundtab}). Most of the projects 
have focused on high-cadence, visible-band photometry of tens of thousands of stars, and only recently
near-infrared filters have started being contemplated. Solar-type stars have been so far the
main focus of all searches, but in recent times the MEarth project (\citealt{nutz08}; 
but see also Damasso et al., this volume)
and the UKIRT WFCAM Transit Survey (WTS) program (star.herts.ac.uk/RoPACS/)
have identified as targets nearby, bright M dwarfs and fainter, more distant
low-mass stars, respectively. Transit discovery programs 
typically achieve photometric precisions of 3-5 mmag. The best-case
performances of $\sim$ 1-2 mmag are mostly obtained with dedicated follow-up programs at 1-2m
class telescopes.

A photometric precision of $\sim3$ mmag is enough to detect Jupiter- and Saturn-sized companions in transit
across the disk of solar-type stars, or $2-4$ $R_\oplus$ planets transiting M dwarfs
(see the definition of $\Delta F/F_0$ above). If the goal becomes the detection of transits
of Earth-sized planets around solar-type primaries, it is necessary to go to space, in order
to achieve $0.0001-0.00001$ mag photometric accuracy. The CoRoT satellite~\citep{baglin09}, 
the Kepler mission (\citealt{borucki09}; Latham et al., this volume), and the PLATO mission, 
currently under study by ESA (\citealt{catala09}; Catala et al., this volume), have been designed
to reach the above performances (we direct the reader to the above contributions for details on
the science). The first transiting Super-Earth of a solar-type star was
recently announced by the CoRoT team (\citealt{leger09}; see also Rouan et al., this volume).

As mentioned above, the spectroscopic characterization of transiting systems
has been carried out primarily from space, by HST and Spitzer. In the future, the prospects
for detailed atmospheric characterization of transiting planets will rely on the James Webb
Space Telescope (JWST), the SPICA satellite, and the proposed THESIS concept. We refer
to the contributions by Clampin, Enya, and Swain in this volume for details on these projects and their potential.

\subsection{The star's the limit}

It is not uncommon to believe that main-sequence, solar-type stars astrophysics is a solved problem,
for practical purposes. In reality, when it comes to transiting planetary
systems, the knowledge of the central star is oftentimes the limit for the accurate determination of
the most sensitive planetary parameters. The precise characterization of transiting planets is
intimately connected to the accurate determination of a large set of stellar properties 
(activity levels, age, rotation, mass, radius, limb darkening, and composition). Some of them
(activity, rotation) can critically limit the possibility to successfully determine the spectroscopic
orbit of the detected planets. Others (mass, radius, age) are strongly model-dependent quantities,
and the correct evaluation of their uncertainties is not trivial (e.g.,~\citealt{brown09}).
Furthermore, precise measurements of the stellar characteristics become increasingly more challenging
for fainter targets.

As discussed in Section~\ref{followup}, follow-up observations of transiting planet candidates
can be very time-consuming (the RV campaign for CoRoT-7 required over 100 spectra, a total of
more than 70 hrs of observing time distributed over 4 months).
For CoRoT, Kepler, and PLATO confirmation via RV measurements
may not even be feasible below a certain radius size, depending on spectral type
(for reference, the semi-amplitude of the RV motion induced by the Earth at 1 AU on the
Sun is $\sim 9$ cm s$^{-1}$, way below the currently best-achievable precision of 50-100 cm s$^{-1}$
with the HARPS spectrograph). Devices for ultra-stable wavelength calibration such as laser combs
can in principle allow to push towards $<10$ cm s$^{-1}$ precision~\citep{li08}, provided the star
cooperates~\citep{walker08,makarov09}.
Achievable SNR for a given host's spectral type make also complicated the problem of atmospheric
characterization of transiting rocky planets via transmission spectroscopic or secondary-eclipse 
observations~\citep{kalt09,deming09}.

Given the above issues, it is clear that bright ($V\leq 12$) stars are the privileged targets
for transit searches. The challenge is then designing a survey
capable of covering large areas of the sky to maximize
the yield of good targets. This is planned for PLATO, thanks to its step \& stare mode (Catala et al., 
this volume) and, in all-sky fashion, by TESS~\citep{deming09}.

\section{The BDT Perspective}

\begin{figure}
\centering
\includegraphics[width=.15\textwidth,angle=90.]{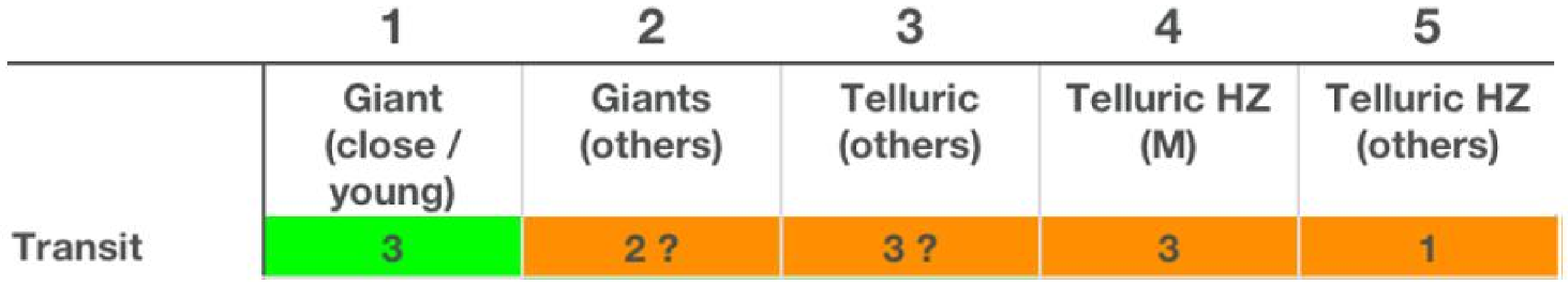}
\caption{Science potential of the family of techniques for transit detection and characterization
for different classes of exoplanets. Number and color coding as described in Coud\'e du Foresto 
et al. (this volume).}
\label{trpot}
\end{figure}

As discussed in detail by Coud\'e du Foresto et al. (this volume), the BDT has devised a strategy
to gauge the interplay between families of techniques, project scale, scientific potential,
and detectable exoplanet class within the context of the multi-step approach recognized necessary
in order to reach the final goal of detection and characterization of terrestrial, habitable planets.
Following the `grid methodology' agreed upon within the BDT, and outlined in detail by
Coud\'e du Foresto et al. (this volume), the TWG has attempted to gauge the potential and
limitations of transit photometry (and follow-up measurements) as a tool to detect
and characterize extrasolar planets, while emphasizing its complementarity with other
techniques. Overall, our preliminary conclusion is that transit-discovery observations
can crucially contribute to statistical studies of planetary systems (science potential 1)
and to identify systems suitable for follow-up (science potential 2). Follow-up observations
of known transiting systems have the potential to achieve their full spectroscopic characterization
(science potential 3). However, the full potential of the technique might not be realized for
all classes of extrasolar planets encompassed in this exercise. In particular (see Figure~\ref{trpot}):

{\bf Hot giant planets}: Ground-based, wide-field transit surveys
 with typical photometric accuracy $<0.01$ mag, have allowed to detect several tens of
 hot Jupiters. The ongoing CoRoT mission is also providing many detections of close-in Giants, and the
 prospects with Kepler are very encouraging too.
 The Spitzer and Hubble Space Telescopes have been utilized as follow-up tools for the
 (broad-band) spectral characterization of several hot Jupiters at visible,
 near, and mid IR  wavelengths, with several molecules identified).

 {\bf Giant planets at large orbital radii}:
 CoRoT and Kepler are capable to achieve an
 accuracy of $10^{-4} - 10^{-5}$ mag, respectively, in the visible.
 They will provide a census of transiting giant planets out to 1 AU based on $\sim10^5$ targets.
 The proposed TESS all-sky survey ($\sim 2012$) could achieve a photometric precision similar to that of CoRoT,
 and could provide a census of transiting giants with periods up to several tens of days around bright stars.
 While not a truly all-sky survey, the PLATO mission ($\sim 2017$)) could achieve, on
 a statistically significant sample of bright stars, a photometric precision exceeding Kepler's.
 It will be sensitive to Jupiters on orbital periods similar to those accessible by Kepler.
 Statistical information on the rate of occurrence of longer-period giant planets will
 also be collected by ongoing and upcoming large-scale ground-based surveys, such as LSST and PANSTARRS
 (see Figure~\ref{groundtab}).
 For the sample of relatively bright stars, several very efficient space- and
 ground-based facilities will become available in the near future for characterizing
 spectroscopically the discovered planets. These include JWST and SPICA (and possibly THESIS)
 for infrared photometry and spectroscopy,
 and the ELTs for high-resolution spectroscopy in the optical and near IR.

 {\bf Telluric Planets in and out of the Habitable Zone of M dwarfs
 and solar-type stars:} CoRoT and TESS have the potential to detect Super-Earth planets around all
 targets, and at a range of orbital radii, including the Habitable Zone of low-mass stars 
 (CoRoT has recently announced its first detection). Kepler has the potential to provide the
 first statistically sound estimate of $\eta_\oplus$. The ultra-high-precision photometry delivered
 by PLATO will also allow the detection of Earth-sized planets in the Habitable Zone of F-G-K targets.
 As for low-mass stars, the ongoing ground-based MEarth cluster of telescopes
 is optimized to search for transiting Super-Earths in the Habitable Zone of nearby M dwarfs,
 while the WTS survey will target a large sample of low-mass stars, searching for
 transiting rocky planets with periods of a few days. 
 Theoretical studies are now maturing, which can predict the range, 
 and strength, of spectral fingerprints of terrestrial, habitable planets 
 (e.g.,~\citealt{grenfell07,kaltetal09}). The proposed SPICA mission, JWST, and the THESIS concept
 will be capable to perform spectral characterization (broad bands, spectra) 
 in the near- and mid-IR possibly down to telluric planets in the HZ of bright M dwarfs.
 The proposed SIMPLE instrument for the E-ELT would also be able to
 perform transmission spectroscopy of low-mass planets transiting M dwarfs.

\section{Summary}

On the `bright' side, transit photometry allows to characterize the bulk composition of a planet,
and it identifies systems suitable for atmospheric characterization. Such studies are
more difficult to undertake for non-transiting systems. In order to fully exploit
the potential of this technique (and follow-up measurements), there does not seem
to be a clear need for a facility devoted to planetary transits which would require
investments on the order of an ESA's flagship (L-class) mission. On the other side, this technique
requires large amounts of follow-up work, and the stellar host can often be the limiting factor
in the precision with which the crucial physical parameters of the planets are determined.

The relevant technology for transit detection of terrestrial-type planets is already available.
Ongoing and future programs have the potential to nail the occurrence
rate of habitable planets around main-sequence stellar hosts, and, provided some degree of
further technological development, characterize those around stars with favorable spectral type.

\acknowledgements 

A.S. gratefully acknowledges financial support from the Italian Space Agency through ASI
contract I/037/08/0 (Gaia Mission - The Italian Participation to DPAC).


\end{document}